\documentclass[aps,prl,preprint,groupedaddress]{revtex4}
\usepackage{graphicx}
\usepackage{color}
\usepackage{amsmath}

\DeclareMathOperator{\atan}{atan}

\begin{document}
%\title[Cosmic Ray contribution to the WMAP polarization]
\title{Cosmic Ray contribution to the WMAP polarization data on the Cosmic Microwave Background}
\author{T. Wibig}
%$^{1,2}$\thanks{E-mail:t.wibigl@gmail.com} }
\affiliation{Faculty of Physics and Applied Informatics, University of \L \'od\'z, ul. Pomorska 149/153, 90-236 \L \'od\'z, Poland}
\affiliation{Astrophysics Division, Cosmic Ray Laboratory, National Centre for Nuclear Research, Poland.}
\author{A. W. Wolfendale}%%$^{3}$%\thanks{a.w.wolfendale@durham.ac.uk}
\affiliation{Department of Physics, Durham University, Durham, DH1 3LE, UK}
%$^{1}$Faculty of Physics and Applied Informatics, University of \L \'od\'z, ul. Pomorska 149/153 90-236 \L \'od\'z, Poland;\\
%$^{2}$Astrophysics Division, Cosmic Ray Laboratory, National Centre for Nuclear Research, Poland.\\
%$^{3}$Department of Physics, Durham University, Durham, DH1 3LE, UK}

\date{}%Accepted 1988 December 15. Received 1988 December 14; in original form 1988 October 11}

%\pagerange{\pageref{firstpage}--\pageref{lastpage}} \pubyear{2002}

%\label{firstpage}

\begin{abstract}
We have updated our analysis of the 9-year WMAP data using the collection of polarization maps looking for the presence of additional evidence for a finite 'cosmic ray foreground' for the CMB.
We have given special attention to high Galactic latitudes, where the recent BICEP2 findings were reported. 
The method of examining the correlation with the observed gamma ray flux proposed in our earlier papers and applied to the polarization data shows that the foreground related to cosmic rays is still observed even at high Galactic altitudes and conclusions about gravitational waves are not yet secure.
Theory has it that there is important information about inflationary gravitational waves in the fine structure of the CMB polarization properties (polarization vector and angle) and it is necessary to examine further the conclusions that can be gained from studies of the CMB maps, in view of the disturbing foreground effects.
\end{abstract}

%\begin{keywords}
%cosmic rays; cosmic microwave background.
%\end{keywords}

\maketitle
\section{Introduction}
The 'structure' of the Cosmic Microwave Background (CMB) is a fertile 
area for the testing of cosmological models, but the elimination
of foreground effects due to processes in the Galaxy, particularly 
in the Halo, is not a trivial problem. The processes relate mainly to
photons due to Cosmic Ray (CR) electrons and radiation from dust. The 
frequency used in the CMB maps (\citet{wmap} hereafter referred to 
as the WMAP) is that for which the foreground contributions are a minimum. It is interesting to note
\citep{banday} that at this frequency the sum of the components gives 
almost the 2.7 K temperature spectral shape.
%}

In our recent publication (\citep{we2-mnras}), which confirmed results reported by us about ten years ago  (\citep{we1-mnras}) with the one-year
only WMAP data, \citet{tegmark}, we have shown that there is strong evidence that the procedure 
of eliminating the known and identified foreground contributions is not perfect.
We have found some contributions 
from large Galactic structures which have not been taken into account yet.

The method we used was based on the correlation of the explicit cosmic ray origin effects measured by the EGRET
satellite \citet{egret}, i.e.   the gamma ray flux, and the CMB temperature characteristics.

In the present work we are still using the EGRET Gamma Ray measurement but comparing the high energy photon flux with the polarization characteristics of the microwave radiation registered by the WMAP experiment.

Some remarks are necessary about the sources of the data. Although the CMB 'temperature' maps used earlier were cleaned  of foreground, to the best degree available, by the authors, 
those for polarization -- as published -- have not been. There is thus an inevitable foreground contribution. Turning to the gamma ray map, the EGRET measurements have been superseded by the Fermi-LAT data. However, a comparison of the two sky maps shows that they are very similar. The features in the maps which are relevant to the present analysis relate largely to the distribution of CR-target material, which is in the form of atomic and (lumpy) molecular hydrogen. It is the interaction of CR with the gas which produces the measured gamma rays. Our hypothesis regarding the CMB foreground is that some, at least, will be due to CR-heated dust and electron synchrotron radiation, both of which are only indirectly connected with the gas. Thus, and in order to preserve continuity with previous work, we continue to adopt the EGRET maps.
% It must also be said that the FERMI-LAT maps are not easy to access. When such easily available gamma ray maps are available it will be useful to repeat the analysis with different gamma ray energy thresholds, the point being that the relative importance of CR electrons and protons varies with energy, and thus the predicted foregrounds will vary somewhat.

\section{The analysis}
The sky map of CMB temperature fluctuations has been analysed in many different ways. The most important and significant is the search for the cosmological structures observed against the background of  random noise. 
On the other hand the observed sky map at any frequency of the electromagnetic spectrum has obvious structures related to the Galactic 'shadow' or to Galactic sources. These structures make a foreground on the CMB sky and we need to identify them and then extract them. 
For the temperature maps the process of the foreground reduction gives the set of "cleaned" maps, which contains to a great extent the real 'cosmological' CMB structures, but the process of cleaning is not obvious and there could be different strategies leading to slightly different results. This is a source of systematics in the further analysis of the CMB data and in turn the source of uncertainties in the resulting cosmological parameters.
%}
For the polarization case the vector character of the data makes the cleaning procedure much more complicated.
In addition, the small degree of polarization makes the flux under consideration much harder to determine experimentally. All this is probably the reason that the set of polarization data of WMAP available to the public do not contain the 'cleaned maps' with any known cleaning procedure.

The analysis of the WMAP polarization data shows
%has to be based on the measurements which are obviously contaminated by the Galaxy, this is clearly seen as 
a strong enhancement of the intensity of radiation close to the
Galactic disk as would be expected for all foregrounds.

The same structure is seen for the gamma ray flux in the EGRET map. We supposed that the EGRET map (with point sources excluded) is a view of the possible foreground on CMB maps caused by the cosmic rays and Galactic matter structures.
The 
correlation between the EGRET gamma ray flux and the temperature fluctuations measured in this direction by WMAP was used in our previous papers \citep{we1-mnras,we2-mnras}. 
We have followed this way of constructing the sky map of the respective correlation coefficients for the polarization
parameters of WMAP.

The polarization is measured as a set of Stoke's parameters ($U$ and $Q$). For each point (pixel) in
the sky the number of independent observations is also given. The statistics (exposure) of 
different parts of the sky is not uniform and there are patterns relevant to the detector structures and measurement technique as
mentioned already in \cite{jarosik}.%Three-Year Wilkinson Microwave Anisotropy Probe (WMAP) Observations: Beam Profiles, Data Processing, Radiometer Characterization and Systematic Error Limits N. Jarosik, et al., 2007, ApJS, 170, 263
This is a factor which contributes slightly to the 
reported average values of polarization. The polarization intensity, $I$, even if there is a completely random polarized light registered, is defined as the square root of $(Q^2+U^2)$, and has non-zero average values. The square of $I$ fluctuates as the sum of the squares
of two randomly distributed quantities. Each value of $U$ and $Q$ is the average value of a number of measurements, so it should in principle obey a normal, Gaussian, distributions with width proportional to the square root of the measurement number. All Stokes parameter values  ($I$, $U$ and $Q$) measured by WMAP depend on how long each fragment of the sky was observed.

This is very important when one is looking for the small structures of polarized CMB.  In the present paper, however, we concentrate on big and very big structures, as were
the ones we found in the temperature fluctuation analysis.

To diminish the effect of random fluctuations we used the technique of the 'running average' thereby smoothing the small scale effects.
The smoothing of the WMAP polarization parameter maps was done by calculating the value of each parameter
for each point on the high resolution WMAP, averaging all the points within a cone of radius 10$^\circ$ weighted with the exponential weight:
$\exp [ {- ({\rm angular\ distance}/ 3 ^\circ })^2]$.

Such smoothed $Q$ and $U$ maps were used to derive the polarization intensity $I$ and the polarization angle
\mbox{$\theta =  1/ 2 \arctan (U/Q)$.}
We define the angle with respect to the N-S direction: value  0 represents the N-S direction and 90$^\circ$ is for the W-E direction. 

\begin{figure}
%  \vspace*{174pt}
\includegraphics[width=12.3cm]{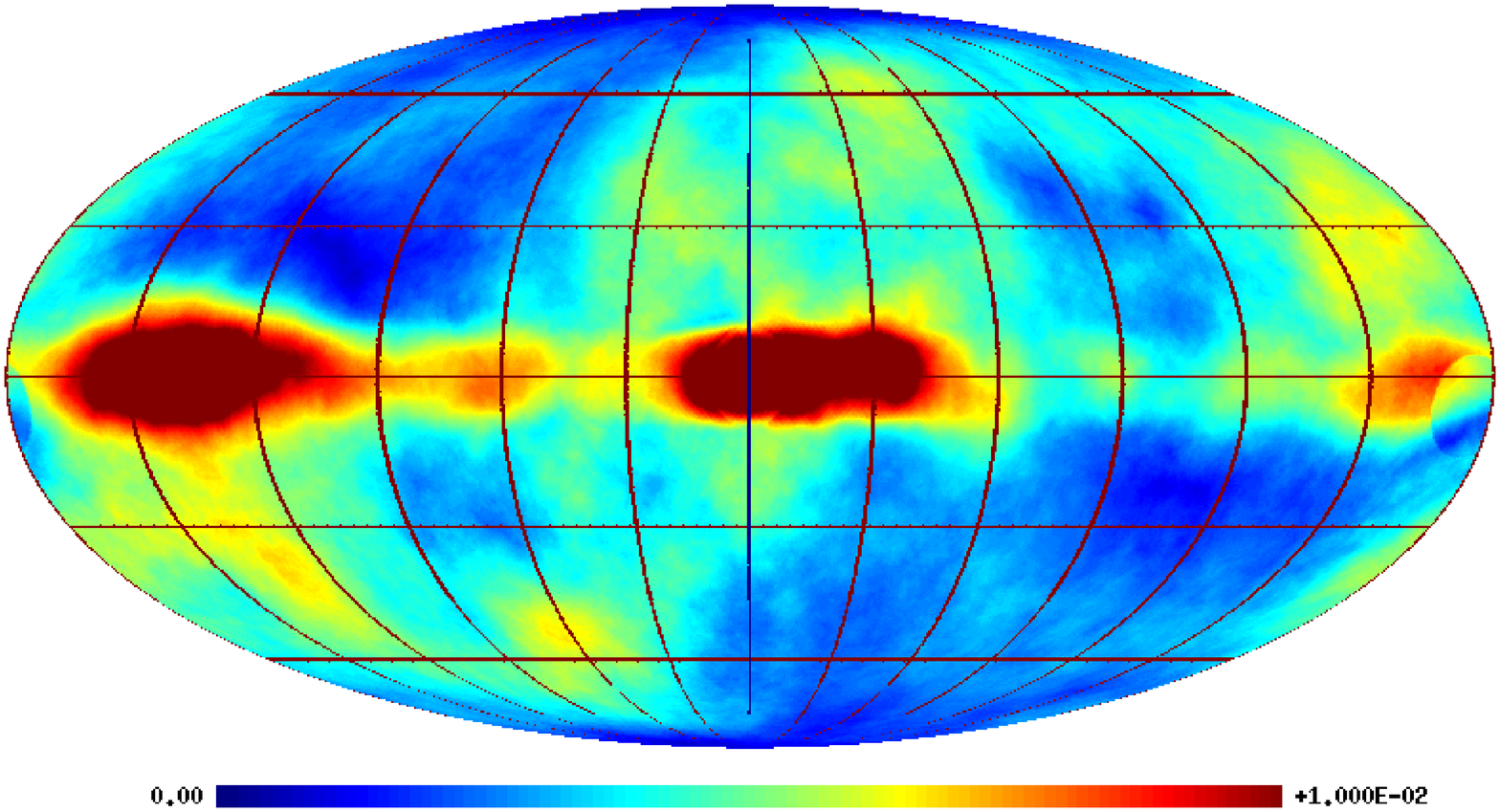} 
\includegraphics[width=12.3cm]{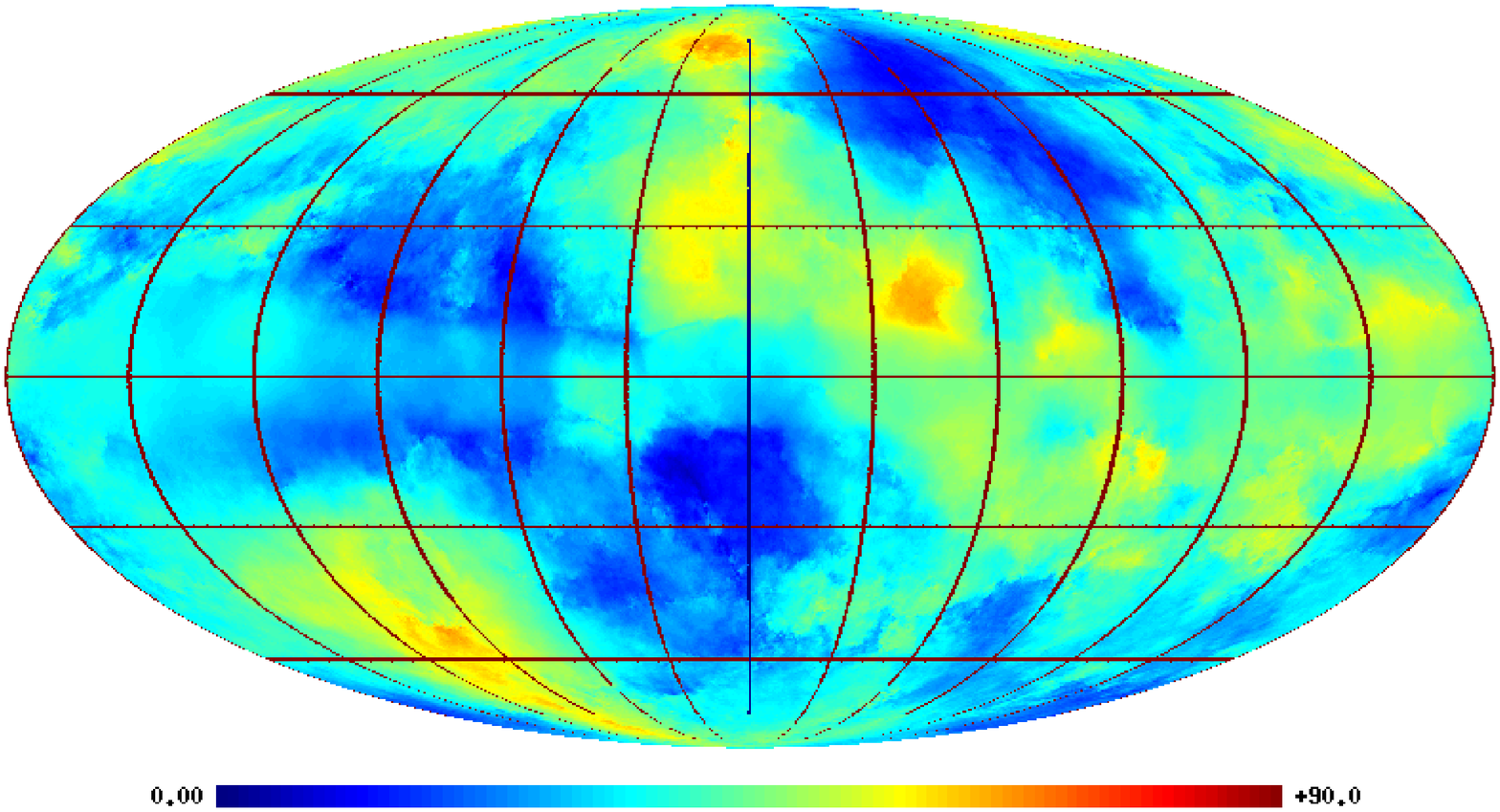} 
  \caption{The WMAP data smoothed (see text) sky maps for polarization (a) and polarization angle (b).
The angle equal to 0 degrees represents the N-S direction while 90$^\circ$ is the W-E direction.  The maps are obtained using the software of the HEALPix \citep{healpix} package.
\label{fig1}}
\end{figure}

The procedure of running average was used to smooth the WMAP data for all frequency bands. Further analysis 
has been performed for each band separately, but also we add the fluxes of Q, V and W bands to produce a
less fluctuating picture. The results and conclusions do not change using the sum, so below we present some
details only for this case. 

The input maps are given in Fig.\ref{fig1}.
 
The polarization map shows the emission from the Galactic disk, which is not excluded from the data sets. There is also visible the Equatorial Plane which could be partially an effect of the statistics discussed above. The angle map does not contain these well defined and understood structures, but in general the angle map could be misleading. At every point the value of \mbox{$\theta =  1/ 2 \atan (U/Q)$} is obtained whatever the values of $Q$ and $U$ are, however small is the total polarization flux itself. Due to the uncertainties of the measurement for small
values of $U$ and $Q$ the polarization direction is highly undetermined (small fluctuations of $Q$ around the value of 0 from a positive to negative value change abruptly the direction from N-S to W-E). The maps for the angle should therefore be used with great care.

\begin{figure}
%  \vspace*{174pt}
\includegraphics[width=12.3cm]{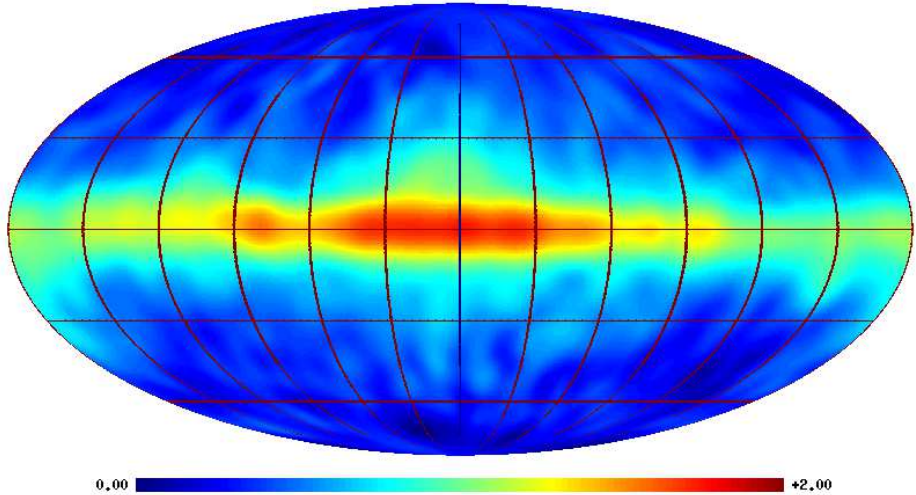} 
  \caption{The EGRET gamma ray intensity map.\label{egr}}
\end{figure}

Maps for both these quantities were then compared with the EGRET intensity map shown in Fig.~\ref{egr}. For each pixel the correlation with gamma ray intensity was calculated as a linear correlation coefficient for all point in the neighbourhood of the pixel defined as the directions in the sky within the cone of radius of 10$^\circ$.

\begin{equation}
\label{c}
c_j={{\langle {\rm E}_j  {\rm W}_j \rangle - \langle {\rm E}_j   \rangle \langle{\rm W}_j \rangle}\over {\sigma_{\rm E}  \sigma_{\rm W}}}~~,
\end{equation} 

where ${\rm E}_j$ is the EGRET gamma ray flux and ${\rm W}_j$ - the polarization or the polarization angle towards the point $j$. %
Parenthesis $\langle ... \rangle$ in Eq.(\ref{c})
 means averaging over all points in the cone centred at the point $j$; $\sigma_{\rm E}$ and $\sigma_{\rm W}$ are respectively the squared variances of the EGRET gamma ray flux and the WMAP polarization signals -- within this cone.

\begin{figure}
%  \vspace*{174pt}
\includegraphics[width=12.3cm]{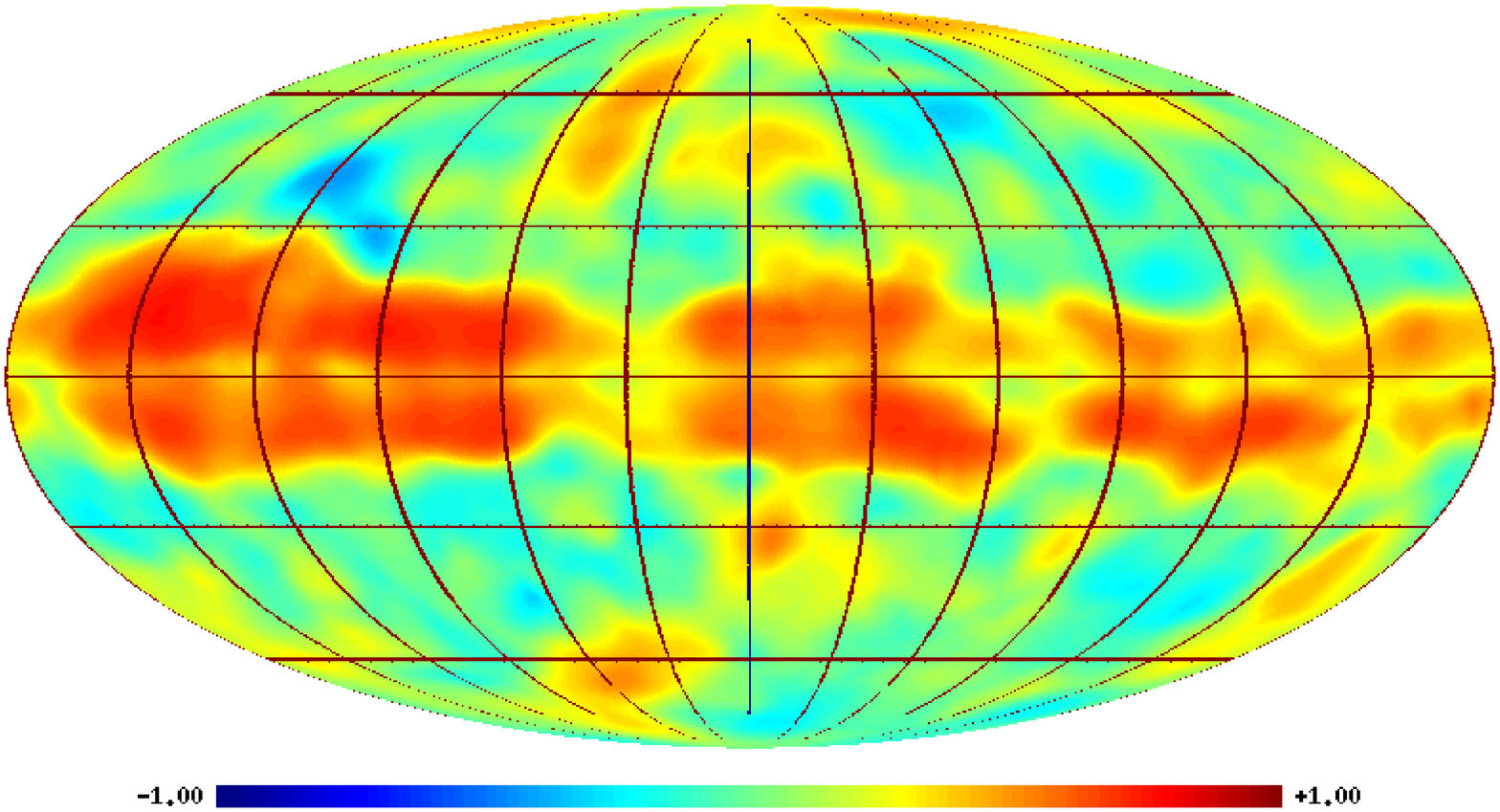} 
\includegraphics[width=12.3cm]{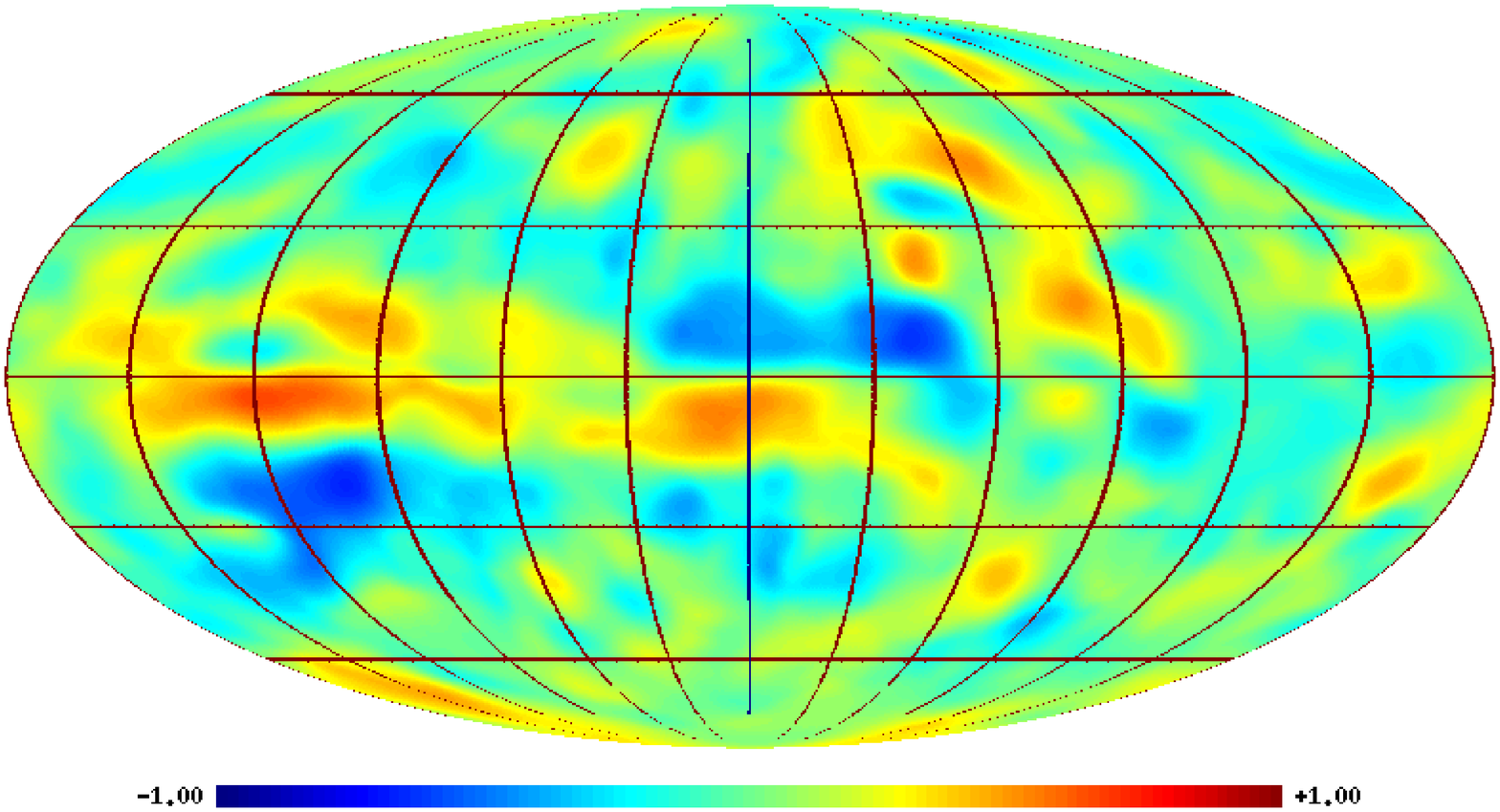} 
  \caption{The WMAP data smoothed (see text) sky maps for polarization (a) and polarization angle (b).
The angle equal to 0 degrees represents the N-S direction while 90$^\circ$ the W-E direction.\label{fig2}}
\end{figure}

The correlation coefficient maps are shown in Fig.~\ref{fig2} for the polarization intensity (a) and the polarization angle (b). The first shows obvious correlation with the disk emission from the EGRET map. The disk component is limited to small $b$ values. Above $b=30^\circ$ it seems to vanish. The map in Fig~\ref{fig2}(b) shows similar patterns limited again to $b<30^\circ$.

In the sky far from the Galactic Plane there are some pattern of quite large sizes seen in Figs.~\ref{fig1} and \ref{fig2}. They could come from random fluctuations ('cosmological variance') or they could be related to some processes determined by big structures in the Universe, most likely in the vicinity of our Galaxy, or some Galactic structures, or even quite local  Galactic structures. 

To study this we have to look  first  at the size spectra of the patterns.   

\subsection{Spherical harmonics}
To examine this structures we can use the spherical harmonic analysis.
It is known that the distribution of the signals on a sphere can be expanded as a linear combination of spherical harmonics:
\begin{equation}\label{sph1}
F(\theta,\varphi)=\sum_{\ell=0}^\infty \sum_{m=-\ell}^\ell f_\ell^m \, Y_\ell^m(\theta,\varphi).
\end{equation}
The effect of filtering (by the smoothing procedure applied to the raw data) was studied in \citet{we2-mnras}. No additional
power resulting from the 'running average' we used is observed in the spectrum for values of  $\ell$ below the respective 
averaging size.

\begin{figure}
\centerline{
\includegraphics[width=8.3cm]{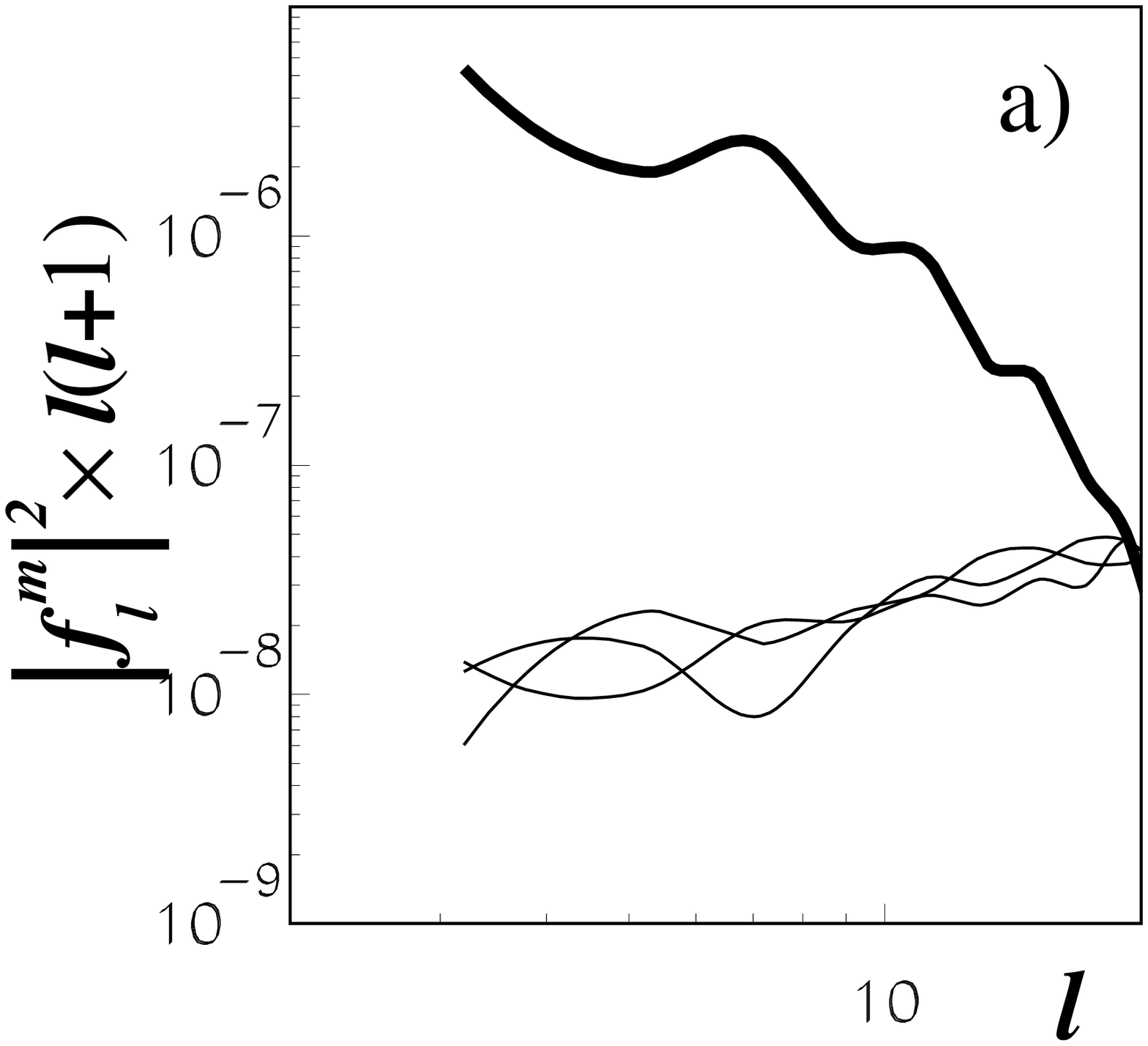} 
\includegraphics[width=8.3cm]{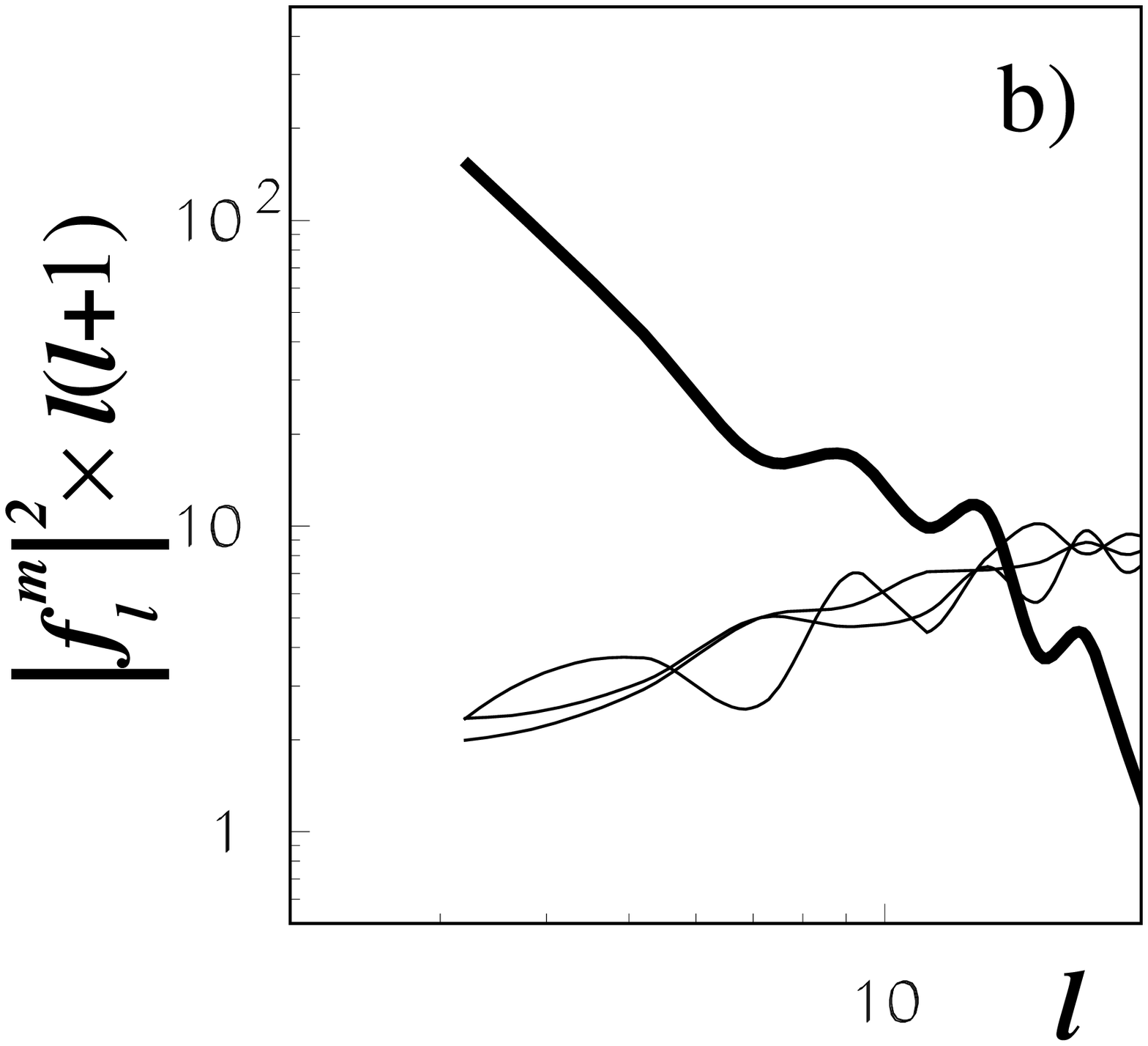} }
  \caption{Smoothed harmonic spectra (shown as $\sum _m |f_\ell^m |^2 \times \ell (\ell+1)$)
calculated for the polarization (a) and polarization angle (b) of the maps shown in Fig.\ref{fig1} (solid lines) compared with the spectra obtained for three examples of 'randomized' skies (see text) spectra . 
\label{powpk}}
\end{figure}

Spectra of the polarization and polarization angles measured by WMAP are shown in Fig.~\ref{powpk}. We compared them with the randomized sky map spectra obtained by shuffling pixels on the original WMAP maps. Such maps do not contain any information about the Galaxy and the position with respect to any real structures, so they can be used as an estimation of the background for our foreground analysis. As we can see in  Fig.~\ref{powpk}, for $\ell$ greater than $20$ the effect of the running average suppressed possible existing forms in the CMB sky. The curious lowering of the angle spectrum even below randomly shuffled pixel sky spectra for $\ell \approx 20$ is not understood: it is a specific feature of the map in Fig.~\ref{fig1}b.

\begin{figure}
\centerline{
\includegraphics[width=8.3cm]{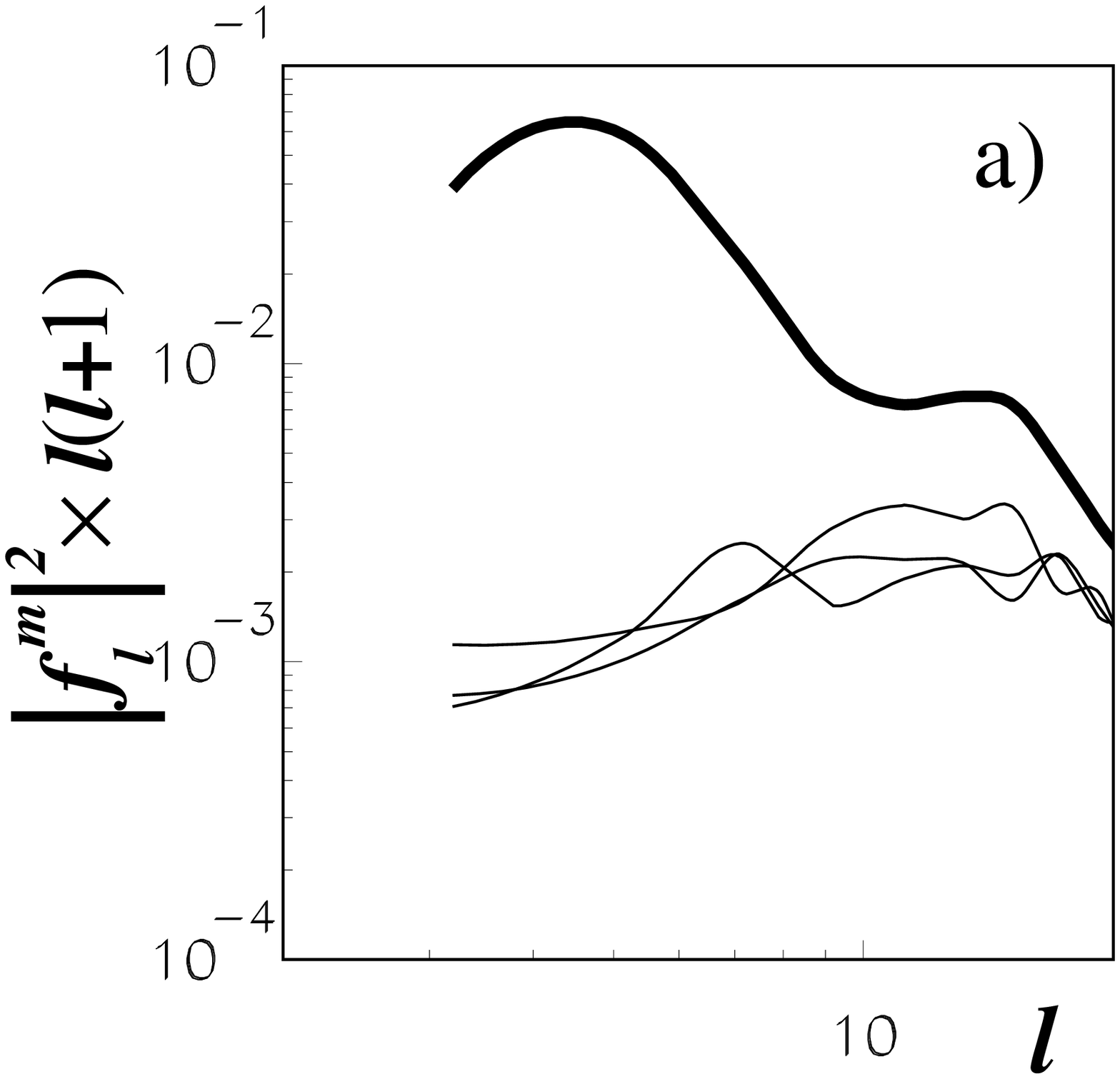} 
\includegraphics[width=8.3cm]{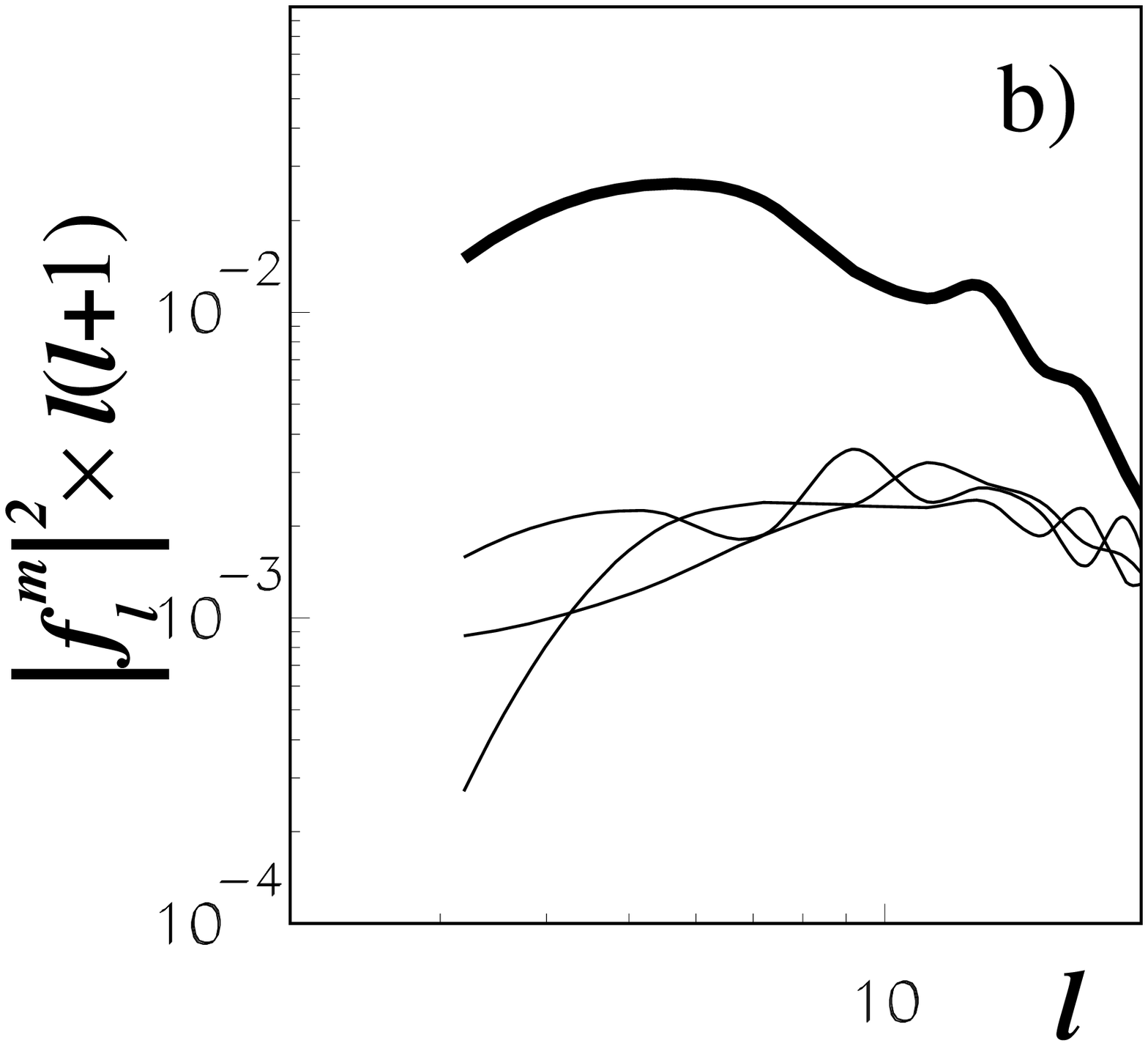} }
  \caption{Smoothed harmonic spectra calculated for the linear correlation coefficient of the EGRET intensity flux and the polarization (a) and the polarization angle (b) of WMAP as shown in Fig.\ref{fig1} represented by the solid lines compared with the spectra obtained for three examples of 'randomized' skies spectra. 
\label{powcor}}
\end{figure}

The correlation coefficients maps suggest the existence of non-random low $\ell$ patterns, too.
The power spectra of maps in Fig.~\ref{fig2}
are shown in Fig.~\ref{powcor}. We can see that the effect of the structures exists in the correlation maps. 
The unusual behaviour of the angle spectra around $\ell = 10 - 20$ has disappeared. Polarization and polarization angle correlation coefficient spectra look quite similar, only for $\ell$ of a few the large Galactic disk related spectral power of the correlations for polarization value is slightly higher than for angle.

To see if the effect is only that related to the well visible Galactic disk, or is extended over high altitudes, we examined the spectra calculated for the parts of the sky with different cuts on Galactic altitude $b$.
A search for more complicated big patterns is almost impossible, mainly because of the uncertainties of the measured values of the polarization light. Instead of plotting the whole spectra, we sum up the values shown in Figs.\ref{powpk} and \ref{powcor} in the low $\ell$ interval:
from 3 to 10. We defined the respective 'spectral power density' as 

\begin{equation}\label{f35}
{\rm f_{3-10}}(b_{\rm cut})={1 \over \Omega(b_{\rm cut})} \sum_{\ell=3}^{10} \sum_{m=-\ell}^\ell |f_\ell^m|^2 \times \ell (\ell+1)~~,
\end{equation}
where $\Omega(b_{\rm cut})$ is the solid angle of the analyzed sky covering the region of $|b|>b_{\rm cut}$.

\section{Results for high altitudes}
\begin{figure}
\centerline{
\includegraphics[width=8.3cm]{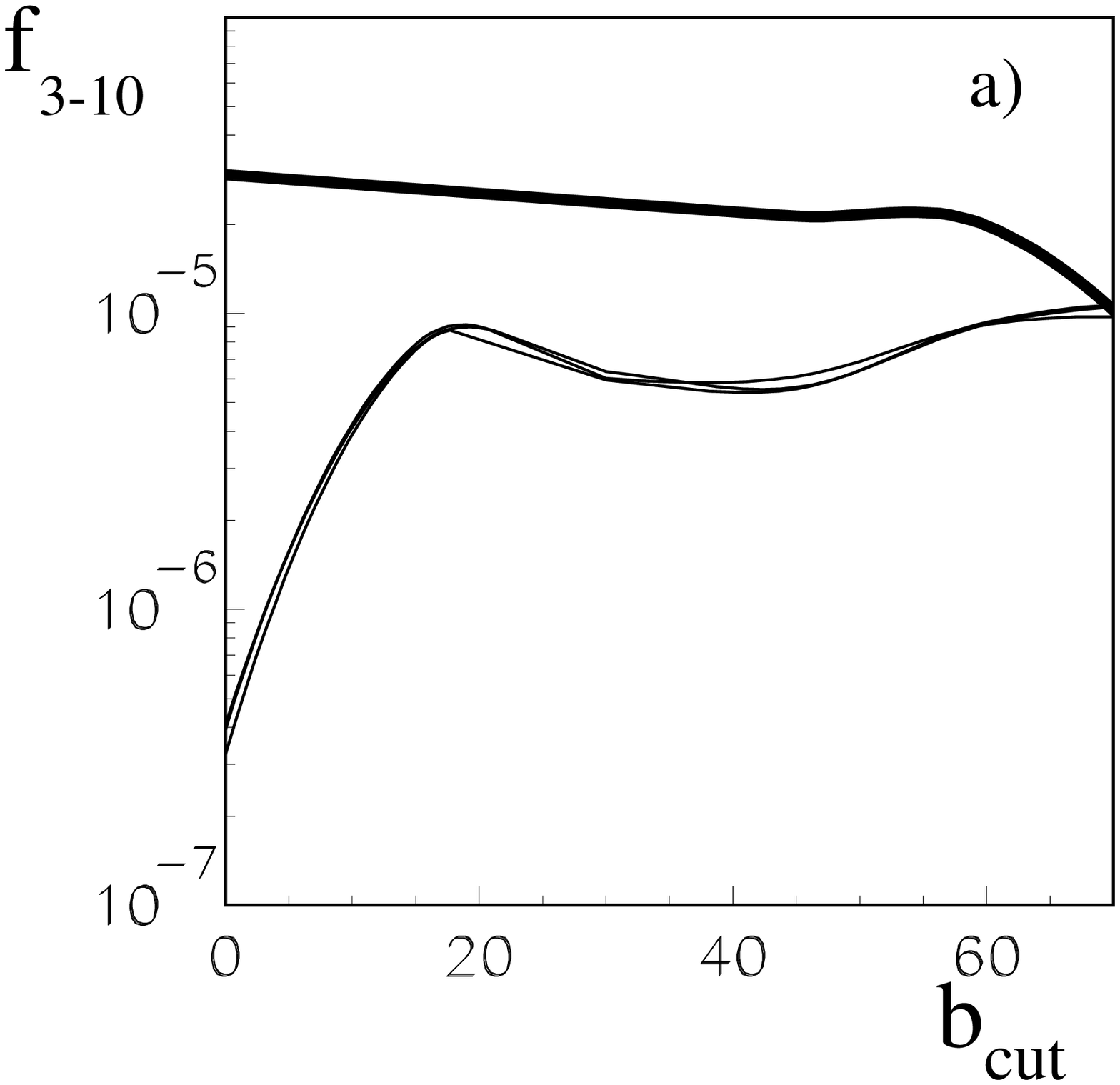} 
\includegraphics[width=8.3cm]{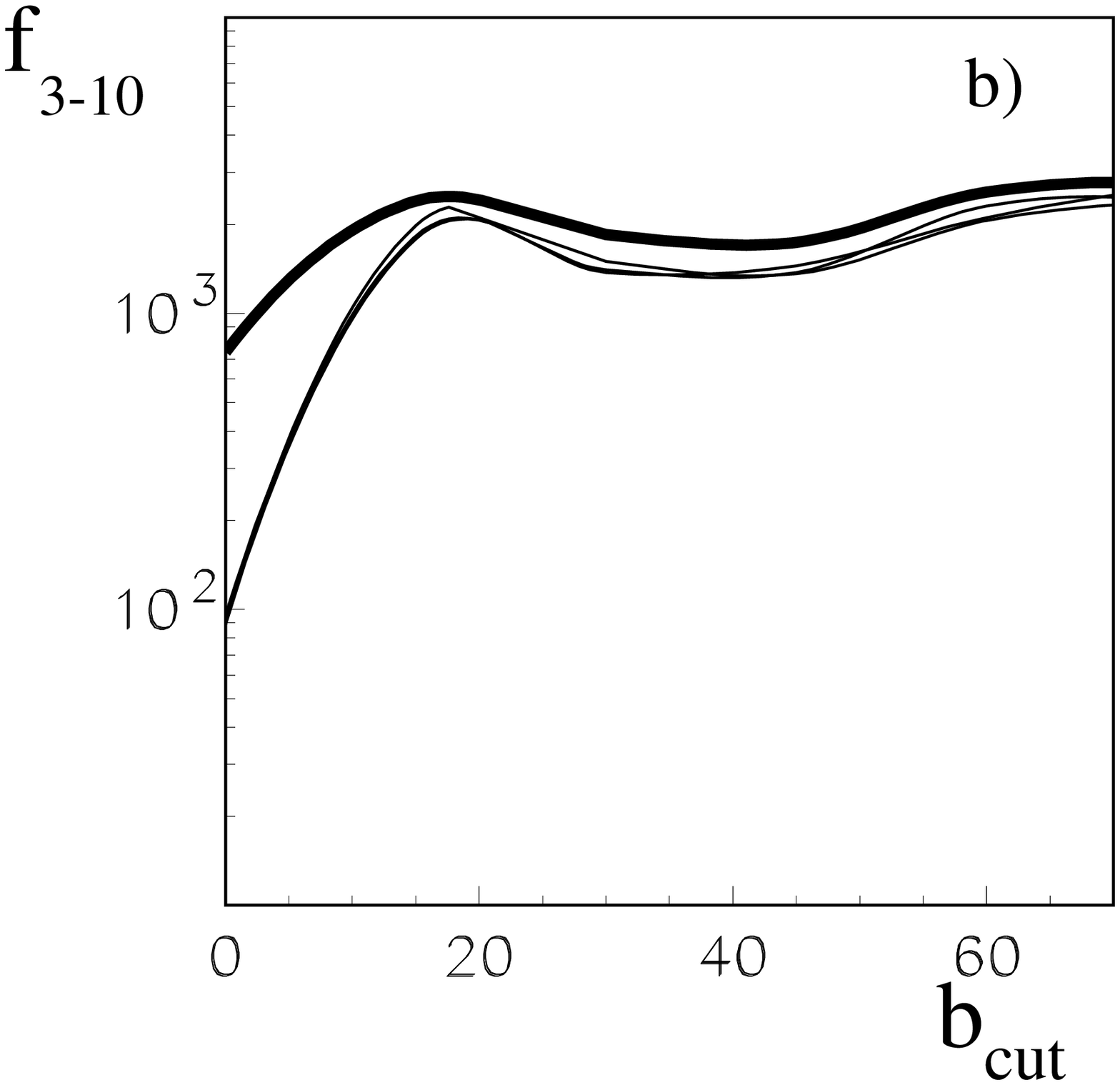} }
  \caption{'Spectral power density' calculated for the polarization (a) and the polarization angle (b), respectively. The results of WMAP, represented by the solid lines, are compared with spectra obtained for three examples of 'randomized' skies spectra.
\label{cutspk}}
\end{figure}

\begin{figure}
\centerline{
\includegraphics[width=8.3cm]{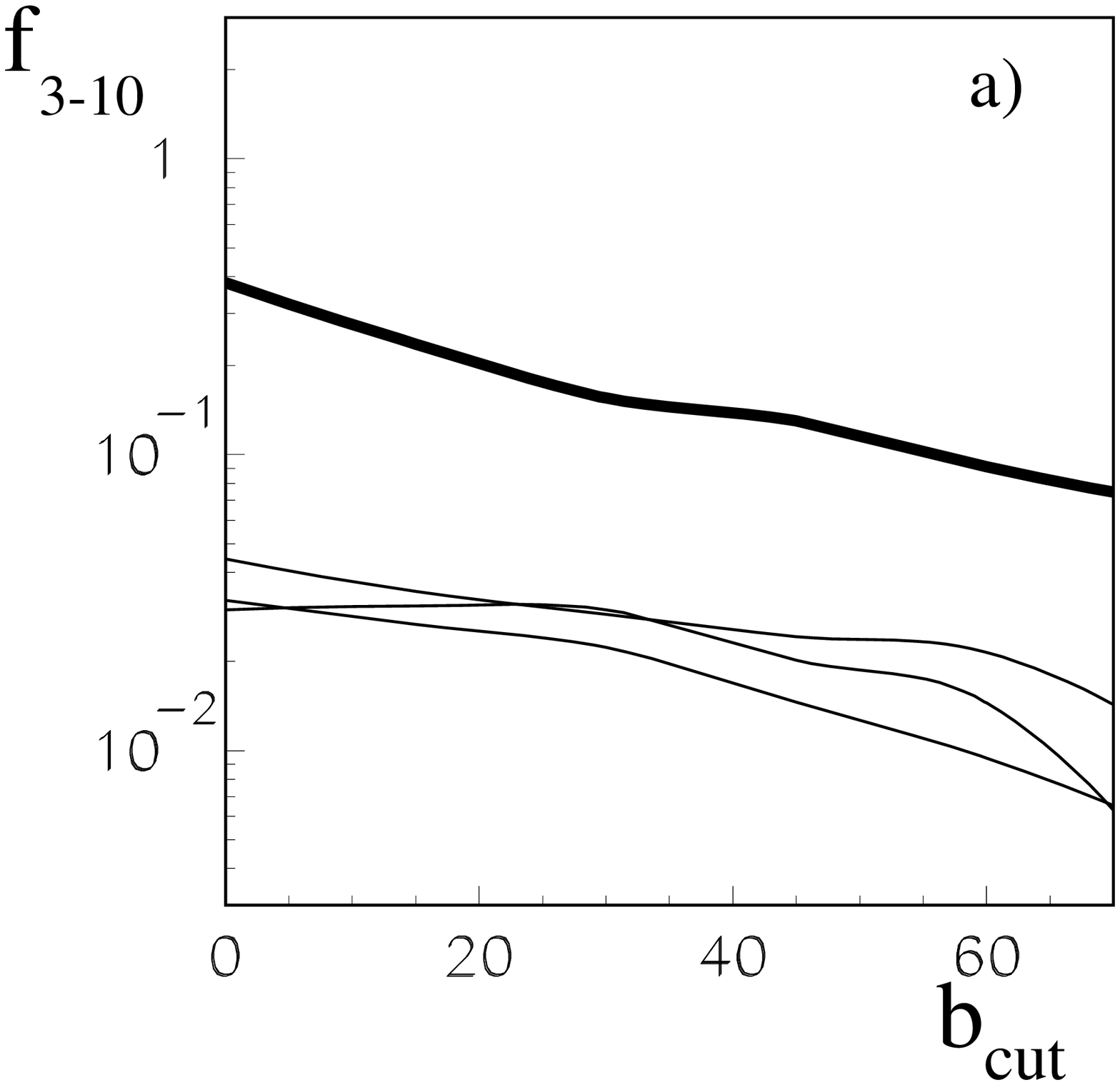} 
\includegraphics[width=8.3cm]{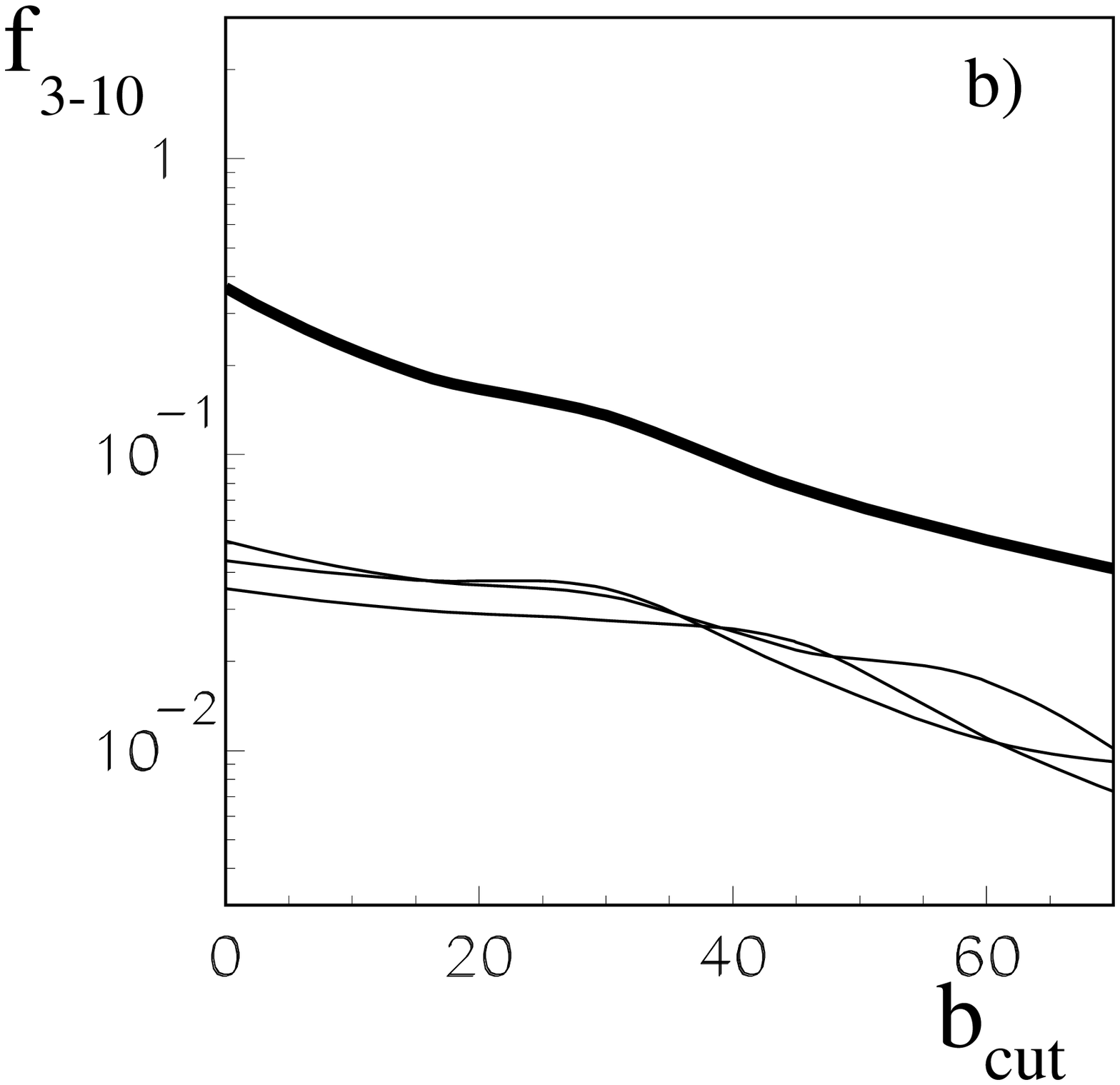} }
  \caption{'Spectral power density'   for the linear correlation coefficient of EGRET flux and the polarization (a) and the polarization angle (b) shown as in Fig.~\ref{cutspk}. The results of WMAP represented by the solid lines compared with the spectra obtained for three examples of 'randomized' skies spectra.
\label{cutscor}}
\end{figure}

Results of our calculations of ${\rm f_{3-10}}$ are shown in Figs.~\ref{cutspk} and \ref{cutscor}. 

As we can see for the polarization value (Fig.~\ref{cutspk}a) there are structures existing to Galactic altitudes of about 60$^\circ$ and these structures are not random patterns. In the case of the polarization angle 
 (Fig.~\ref{cutspk}b) there is an enhancement of the power very close to the Galactic disk, and then the spectral power for low $\ell$ goes close to the random sky spectral power, but is still about 20\% higher. We would like to 
examine if this is by chance, or is a real effect. 

This question can be answered, if we check if these structures are related to the Galactic structures observed
 in the gamma ray intensity map of the EGRET experiment. The respective sums of power in the low $\ell$ parts of the linear correlation coefficient for polarization and polarization angle
are shown in Figs. \ref{cutspk} and \ref{cutscor}.

Comparing the power contained in the correlation coefficient spectra three features are seen:
\begin{itemize}
\item[-] the low $\ell$ spectral power is observed both in the polarization intensity and in the polarization angle,
\item[-] in both cases the $b_{\rm cut}$ dependence is very similar
\item[-] the existence of the non-random structures is observed in the whole sky, up to $b=75^\circ$.
\end{itemize}
This could stand for evidence that the polarization intensity and the angle are related to the gamma ray flux rather than to the Galactic structure itself. 

\begin{figure}
%  \vspace*{174pt}
\includegraphics[width=12.3cm]{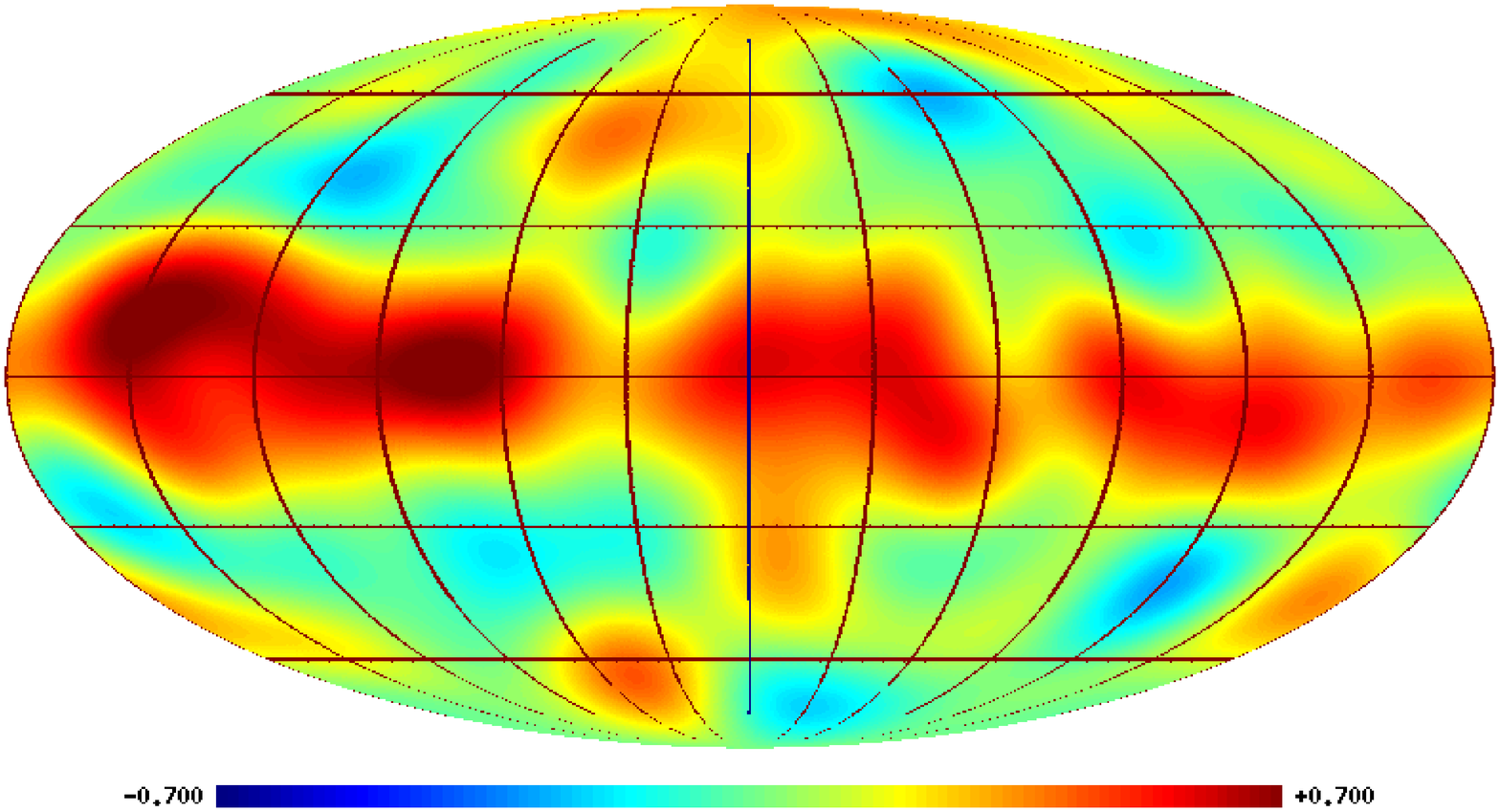} 
\includegraphics[width=12.3cm]{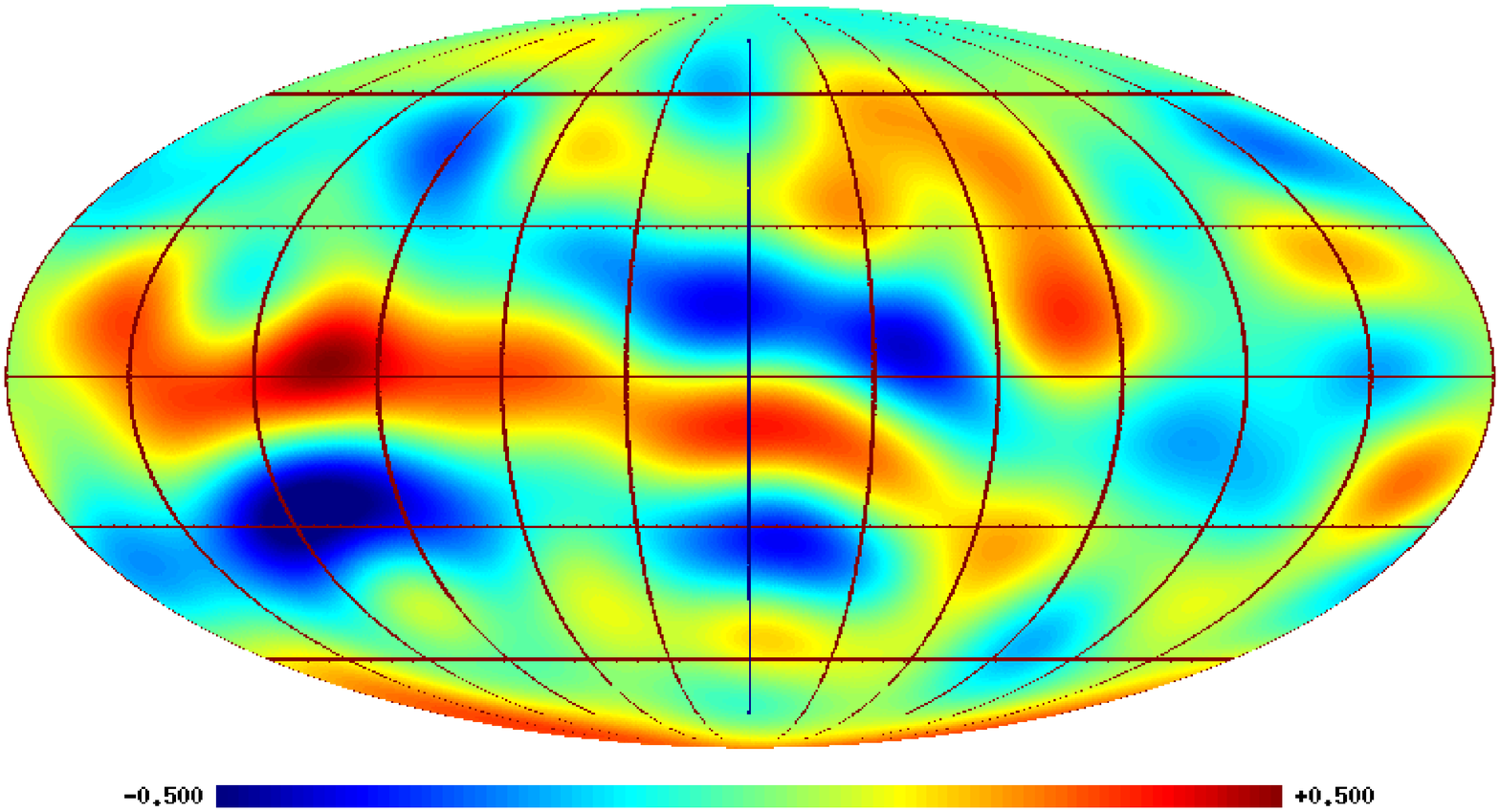} 
  \caption{
Sum of low $\ell$ ($<10$) spherical harmonics for the correlation map of the polarization (a) and the polarization
angle (b) .
 \label{low}}
\end{figure}

We plot in Fig.~\ref{low} the sum of 10 of the first $\ell$ harmonic spectral components for the linear correlation coefficient of the EGRET flux and the polarization value and the polarization angle. The maps consists of a number of blobs of size  $20^\circ - 30^\circ$ which are distributed all over the high $b$ regions. 

\section{Conclusions}
We have examined the CMB WMAPs in some detail from the standpoint of searching for Galactic foreground effects. They appear to be there, at all Galactic latitudes and not just near the Galactic Plane.This conclusion follows from a comparison of the polarization  maps with those of Cosmic Gamma Rays. The presence of a significant foreground in the latitude range  $-50$ to $-65$ degrees, is such as to render the interpretation of the BICEP-02 polarization data  \citep{bicep} %( Ade et al., 2014) 
in terms of inflationary gravitational waves untenable at the present time (see, e.g., \citet{Adam:2014bub}). However, the fact that the foreground does diminish to some extent with increasing latitude, $|b|$, suggests that a more sophisticated analysis might be trustworthy.

\section*{acknowledgements}
{The Kohn Foundation is thanked for supporting this work.}

\label{lastpage}


\begin{thebibliography}{99}
%\bibitem[\protect\citeauthoryear{Abassi et al.}{2014}]{ta}Abbasi, R.U. et al., 2014, ApJ 790 L21.
\bibitem[\protect\citeauthoryear{Adam et al.}{2014}]{Adam:2014bub}
  R.~Adam {\it et al.}  [Planck Collaboration],
  %``Planck intermediate results. XXX. The angular power spectrum of polarized dust emission at intermediate and high Galactic latitudes,''
  arXiv:1409.5738 [astro-ph.CO].
   \bibitem[\protect\citeauthoryear{Ade et al.}{2014}]{bicep} Ade, P. A. R. et al.,,2014, PRL, 112, 241101.
  \bibitem[\protect\citeauthoryear{Banday \& Wolfendale}{1991}]{banday} Banday, A. J., Wolfendale, A. W., 1991, MNRAS, 252, 452.
  \bibitem[\protect\citeauthoryear{Bennett et.al.}{2013}]{wmap}%Nine-Year Wilkinson Microwave Anisotropy Probe (WMAP) Observations: Final Maps and Results
Bennett, C. L., et.al., 2013, ApJS., 208, 20B.
   \bibitem[\protect\citeauthoryear{Esposito et al.}{1999}]{egret}Esposito, J. A. et al. 1999 ApJS 123 203;\\
http://heasarc.gsfc.nasa.gov/docs/journal/cgro7.html. 
\bibitem[\protect\citeauthoryear{Gorski et.al.}{2005}]{healpix}Gorski, K. M., Hivon, E., Banday, A. J., Wandelt, B. D., Hansen,
F. K., Reinecke, M., \& Bartlemann, M. 2005, ApJ, 622, 759.

\bibitem[\protect\citeauthoryear{Jarosik et al.}{2007}] {jarosik}Jarosik, N., et al., 2007, ApJS, 170 263.
\bibitem[\protect\citeauthoryear{Tegmark, de Oliviera-Costa \& Hamilton}{2004}] {tegmark} Tegmark, M., de Oliviera-Costa, A., Hamilton, A., 2003, Phys. ReV., D, 68 ,123523. 
   \bibitem[\protect\citeauthoryear{Wibig \& Wolfendale}{2005}]{we1-mnras} Wibig T. and Wolfendale, A. W., 2005, MNRAS, 360, 236.
   \bibitem[\protect\citeauthoryear{Wibig \& Wolfendale}{2015}]{we2-mnras} Wibig T. and Wolfendale, A. W., 2015, MNRAS, 448, 1030.
\end{thebibliography}
\end{document}